\documentclass[12pt,twocolumn,a4paper]{aastex63}
\hypersetup{linkcolor=cyan,citecolor=blue,urlcolor=cyan}
\usepackage{amsmath}

\usepackage{xspace}

\newcommand{\cdig}{C$_{DIG}$\xspace}

\newcommand{\Q}{ionization parameter\xspace}

\newcommand{\hii}{H\,\textsc{II}\xspace}

\newcommand{\Z}{metallicity\xspace}

\newcommand{\doha}{$\rm \Delta log[O\textsc{I}]/H\alpha$\xspace}

\newcommand{\sfrfuv}{SFR(UV$\rm _{corr}$)\xspace}
\newcommand{\sfrha}{SFR(H$\rm \alpha_{corr}$)\xspace}

\usepackage{float}
\usepackage[caption = false]{subfig}
\usepackage{xcolor}

\shorttitle{Spatially resolved comparison of SFRs from UV and H$\alpha$}
\shortauthors{Tomi\v{c}i\'{c} et al.}

\graphicspath{{./}{figures/}}

\begin{document}

\title{Spatially resolved comparison of SFRs from UV and H$\alpha$ in GASP gas-stripped galaxies}

\correspondingauthor{Neven Tomi\v{c}i\'{c}}
\email{neven.tomicic@inaf.it}

\author[0000-0002-8238-9210]{Neven Tomi\v{c}i\'{c}}
\affiliation{Department of Physics, Faculty of Science, University of Zagreb, Bijeni\v{c}ka 32, 10 000 Zagreb, Croatia}

\author[0000-0002-4382-8081]{  Ariel Werle}
\affiliation{INAF- Osservatorio Astronomico di Padova, Vicolo Osservatorio 5, 35122 Padova, Italy}

\author[0000-0003-0980-1499]{Benedetta Vulcani}
\affiliation{INAF- Osservatorio Astronomico di Padova, Vicolo Osservatorio 5, 35122 Padova, Italy}

\author[0000-0003-1581-0092]{ Alessandro Ignesti}
\affiliation{INAF- Osservatorio Astronomico di Padova, Vicolo Osservatorio 5, 35122 Padova, Italy}

\author[0000-0002-1688-482X]{ Alessia Moretti}
\affiliation{INAF- Osservatorio Astronomico di Padova, Vicolo Osservatorio 5, 35122 Padova, Italy}

\author[0000-0001-5840-9835]{Anna Wolter}
\affiliation{INAF - Osservatorio Astronomico di Brera, via Brera, 28, 20121, Milano, Italy}

\author[0000-0002-1734-8455]{Koshy George}
\affiliation{Faculty of Physics, Ludwig-Maximilians-Universit{\"a}t, Scheinerstr. 1, Munich, 81679, Germany}

\author[0000-0001-8751-8360]{Bianca M. Poggianti}
\affiliation{INAF- Osservatorio Astronomico di Padova, Vicolo Osservatorio 5, 35122 Padova, Italy}

\author[0000-0002-7296-9780]{Marco Gullieuszik  }
\affiliation{INAF- Osservatorio Astronomico di Padova, Vicolo Osservatorio 5, 35122 Padova, Italy}

\begin{abstract}

Star-formation rates (SFR) in galaxies offer a view of various physical processes across them, and are measured using various tracers, such as  H$\alpha$ and UV. 
Different physical mechanisms can affect H$\alpha$ and UV emission, resulting in a discrepancy in the corresponding SFR estimates ($\Delta SFR$). 
We investigate the effects of ram pressure on the SFR measurements and $\Delta SFR$ across 5 galaxies from the GASP survey caught in the late stages of gas stripping due to ram pressure. 
We probe spatially resolved $\Delta SFR$ at pixel scales of 0.5 kpc, and compare disks to tails, and regions dominated by the dense gas to diffuse ionized gas (DIG) regions.
The regions dominated by dense gas show similar SFR values for UV and H$\alpha$ tracers, while the regions dominated by the DIG  show up to 0.5 dex higher SFR(UV). 
There is a large galaxy-by-galaxy variation in $\Delta SFR$, with no difference between the disks and the tails. 
We discuss the potential causes of variations in $\Delta SFR$ between the dense gas and DIG areas. 
 We conclude that the dominant cause of discrepancy are recent variations in star formation histories, where star formation recently dropped in the DIG-dominated regions leading to changes in $\Delta SFR$.
The areal coverage of the tracers show areas with H$\alpha$ and no UV emission; 
these areas have  LINER-like emission (excess in $\rm [O\textsc{I}\lambda6300]/H\alpha$ line ratio), indicating that they are ionized by processes other than star-formation.

\end{abstract}


%\maketitle

%%%%%%%%%%%%%%%%%%%%%%%%%%%
%%%%
%%   INTRODUCTION
%%%%
%%%%%%%%%%%%%%%%%%%%%%%%%%%

\section{Introduction} \label{sec:intro}

 The process of star formation, SF (\citealt{Schmidt59, Kennicutt98}), profoundly affects the evolution of galaxies and the physics of the interstellar medium (ISM), but it is also affected by physical processes such as stellar feedback (\citealt{Leroy08, Krumholz09}, \citealt{Barnes21}), magnetic forces (\citealt{Federrath12, Federrath15}), ram-pressure due to inter-cluster medium (ICM; \citealt{GunnGott72, Poggianti17, Vulcani18c, Lize21}, \citealt{Boselli22}), gravitational tidal forces (\citealt{Larson78, Lonsdale84, Renaud14, Tomicic18, Renaud22}), (\citealt{Utomo18, Murphy22}).      
Therefore, a properly measured star formation rate (SFR) across galaxies is an important property used for determining galaxy evolution and its correlation with various physical processes. 
The most common method for precisely measuring SFRs is to use prescriptions that convert measured multi-wavelength tracers into SFR values (\citealt{Leitherer99, KennicuttEvans12, Thilker07}). 
These prescriptions were estimated partly empirically and partly theoretically, exploring the connection between the SFR and the tracers of emission from various ISM components (\citealt{Leitherer99, Calzetti07, KennicuttEvans12, Murphy11, IglesiasParamo06, Tomicic19}). 

The two most often used tracers of current SF are hydrogen Balmer emission lines (more commonly H$\alpha$)  and the ultra-violet (UV) continuum emission.    
H$\alpha$ is emitted by recombining ionized Hydrogen. 
In star-forming regions (HII regions) the gas is ionized by young, massive stars (O type stars with a lifespan of $ \leq7\,$Myr and masses of $\rm M_\star   \geq20\, M_\odot$, and B type stars with a lifespan of $ \leq320\,$Myr; \citealt{Meurer09}) in the OB associations.  
Due to the short lifespan of the O stars, H$\alpha$ typically lasts for no more than 10\,Myr, thus it traces the most recent SF.  
The UV emission instead comes as a direct light emitted from newly formed, but less massive stars (including O type stars with masses $\rm M_\star   \geq3\, M_\odot$; \citealt{Meurer09}) with a lifespan longer than 7\,Myr and up to 200\,Myr. 
This leads to UV tracing not only the most recent SF, but also the phase of SF history older than the H$\alpha$ phase.  

To probe the effects of physical processes across galaxies, it is worth examining the difference in SFRs between the extinction corrected  H$\alpha$ ($\rm H\alpha_{corr}$) and UV ($\rm UV_{corr}$) tracers.  
Thus, we define a difference in the SFR values in this paper as:

\begin{equation}
 \rm   \Delta SFR\,=\,log_{10}SFR(UV_{corr})-log_{10}SFR(H\alpha_{corr}). 
\label{eq:Del sfr}
\end{equation}

Normal star-forming, spiral galaxies at integrated galactic scales should yield negligible $\rm \Delta SFR$.
These spatial scales probe continuous star-formation and the entire range of stellar populations, and therefore the empirical estimation and analytical modeling of the SFR prescriptions for the tracers yield the same SFRs, and $\rm \Delta SFR\approx 0$. 
Sub-galactic $\rm \Delta SFR$ would differ in the case of changes in the relative emission of the tracers due to some physical process. 
Such physical processes could be:  tidal forces, change in SFR efficiency, additional emission from the diffuse ionized gas, stellar feedback, change in stellar composition, change in a relative dust/gas/star distribution, etc. (\citealt{ Calzetti94, Paramo04, Lee09, KennicuttEvans12}, \citealt{Lee16}, \citealt{Tomicic18,  Tomicic19, Moretti20}). 
Some observational and theoretical uncertainties would also yield changes in $\rm \Delta SFR$, such as effects of attenuation measurements, assumptions of SFR prescriptions, changes in spatial resolution, duration of observation,   etc. 

For example, positive values in  $\rm \Delta SFR$ were observed in M33 ($\approx 0.2$dex; \citealt{Verley09}) and in low-mass galaxies of the local universe, with \sfrfuv being higher than  \sfrha (\citealt{Sullivan00, Bell01, Salim07}).
\citet{Sullivan04} and \citet{Paramo04} claimed that the time scales of SF in normal galaxies (instantaneous vs. continuous SF) and in starbursty, low-mass galaxies (which increase the fraction of lower mass and longer lived stars that emit in UV) play a significant role in  $\rm \Delta SFR$ variations.
Another source of variations in  $\rm \Delta SFR$  may be a different shape of the stellar initial mass function (IMF), which may lack a population of OB stars (\citealt{Meurer09, PflammAltenburg09}).  
This might be caused by different physical environments affecting the SF and IMF, or 
by stochastic effects of under-sampling IMF (\citealt{Thilker07, Boissier07}, \citealt{Lee09b,  Lee16}).
Variations in $\rm \Delta SFR$ may also be caused by systematic uncertainties in measuring \sfrha or \sfrfuv. 
A large fraction of the diffuse ionized gas (DIG; \citealt{Walterbos94, Haffner09, Pedrini22})  in the  H$\alpha$ emission of galaxies can increase values of \sfrha compared to \sfrfuv. 
The DIG is a more diffuse gas component compared to the star-forming regions, and spans large spatial scales ($>$kpc) with different relative distributions of gas, dust, and stars. 
This relative distribution may also alter the assumption about the attenuation curve and measurements of photon absorption (\citealt{Tomicic17}), thus altering measurements of SFRs and $\rm \Delta SFR$ (\citealt{Gonzales02}). 
Additional uncertainty in \sfrha may be caused by the fact that the DIG's source of ionization might not be dominated by the SF process alone (\citealt{Poggianti19b, Tomicic21b, Campitiello21, Vollmer21, Pedrini22, Sun22}).

One environment where the diffuse gas component dominates across galaxies is the case of gas-stripped galaxies (\citealt{Tomicic21a}, \citealt{Tomicic21b}), which are in-falling into galactic clusters (\citealt{GunnGott72}).  
The in-falling galaxies show long, stripped ionized gas tails due to the effects of the ICM exerting a ram-pressure on their galactic ISM. 
To observe the effects of ram-pressure on the ISM and galaxies, it is interesting to probe its effects on SFRs and  $\rm \Delta SFR$ across their disks and stripped tails, and in dense gas and DIG-dominated regions\footnote{The DIG-dominated regions are those whose DIG fraction in  H$\alpha$ emission is $\rm C_{DIG}>30\%$. The DIG fraction, $\rm C_{DIG}$ is measured in \citet{Tomicic21a}.}. 
The attempts to observe and compare H$\alpha$ and UV tracers in gas-stripped galaxies were previously done by various groups (\citealt{Gavazzi01, Abramson11, Fumagalli11, Boselli18, Smith10, Vollmer21, Junais21, Boselli21,  Laudari22, George18, Rampazzo22, George23, Gullieuszik23}), but it was rarely done with the comparison between \sfrha and \sfrfuv in their tails, and between star-forming spiral regions and DIG dominated regions in the tails at the same time. 	

The previous attempts of measuring $\rm \Delta SFR$ in integrated star-forming clumps of a few strongly stripped galaxies (referred to as Jellyfish galaxy) were done by \citet{George18}, \citet{George19}, \citet{Poggianti19b}, and \citet{George23} as a part of the GASP project (GAs Stripping Phenomena in galaxies\footnote{https://web.oapd.inaf.it/gasp/}; \citealt{Poggianti17}).
The GASP survey is a multi-wavelength survey that studies gas-stripping processes in 114 galaxies in clusters (the  WINGS and OMEGAWINGS catalogs; \citealt{Fasano06}, \citealt{Gullieuszik15}) and in the field (the PM2GC catalog; \citealt{Calvi11}). 
However, all these observations did not quantitatively measure the spatially resolved, pixel-by-pixel $\rm \Delta SFR$, nor did they compare the dense gas and DIG-dominated areas.    

This work is the first pixel-by-pixel comparison in UV and H$\alpha$ for the GASP survey, and expands the investigation to  DIG-dominated regions. 
The main goals of this paper are: 1) to measure pixel-by-pixel variations of  $\rm \Delta SFR$ in disks and stripped tails, 2) to compare $\rm \Delta SFR$ in dense gas and DIG dominated pixels, and 3) to investigate potential physical sources of variations in $\rm \Delta SFR$ between different regions, such as effects of SFR prescriptions and attenuation curves, changes in escape fractions of ionizing photons, sampling of the IMF, and time scales of SF.    

The paper is organized as follows: the galaxy sample, and observed optical (H$\alpha$) and UV data are presented in Sec. \ref{sec:Data},  the results, which include the analysis of the  $\rm \Delta SFR$ maps, are presented in Sec. \ref{sec:Results}.
In Sec. \ref{sec:Discussion} we discuss all the potential causes of the variation in $\rm \Delta SFR$ and we conclude in Sec. 5. 
In this paper we adopted standard cosmological constants of H$_0=70$ km$\rm \,s^{-1}Mpc^{-1}$,  $\rm \Omega_M=0.3$, $\Omega_\Lambda=0.7$, and the initial mass function (IMF) from \citet{Chabrier03}.

%%%%%%%%%%%%%%%%%%%%%%%%%%%
%%%%
%%   DATA
%%%%
%%%%%%%%%%%%%%%%%%%%%%%%%%%

\section{Data} \label{sec:Data}

%%%%
%%   Subsec
\subsection{Galaxy sample }\label{subsec:Data, gasp, obs}

In this paper, we probe SFR tracers of five gas-stripped galaxies in the GASP survey, which were observed both by optical and UV-based telescopes. 
These galaxies are JO201, JW100, JO60,  JW39, and JO194 (Tab. \ref{tab:Tab01}), and they represent a sample of gas-stipped galaxies with well-defined long tails of ionized gas and recent star formation outside their disks. 
Their redshifts, position on the sky, stellar mass and SFR can be found in \citet{Vulcani18b}, and belong to the optical  WINGS/OMEGAWINGS catalogs of galaxy clusters (\citealt{Fasano06, Gullieuszik15}). 
They are chosen as clear representatives of galaxies undergoing ram-pressure, due to exhibiting the longest stripped tails both in ionized gas and in young stars (visible in the optical and UV) in the GASP sample, and showing cases of unwinding spiral arms (\citealt{Bellhouse21}).  
We note that JO60 and JW100 have a high inclination, with JW100 being edge-on.
\citet{Vulcani20b} studied the spatially revolved SFR for all these galaxies.

\begin{table*}[th!]
\centering
\caption{Names,  galaxy cluster, right ascension (RA.) and declination (DEC.), redshift (z), inclination, and UVIT data availability of the observed galaxies.}
\begin{tabular}{lllllll} \hline \hline
Name &  Galaxy cluster & RA. (deg.) & DEC. (deg.) & z & incl. ($^\circ$) & UVIT data \\
\hline 
JO201 &  Abell 85  & 10.388208 & -9.273028 & 0.0446 &  41.8 & FUV \& NUV  \\
JO60 & Abell 1991  & 223.464875 & 18.651767 & 0.062187 & 69.5 &  FUV \& NUV \\
JW100 & Abell 2626  & 354.104416 & 21.1507 & 0.06189 & 75.1 & FUV \& NUV  \\
JW39 & Abell 1668  & 196.032125 & 19.2106905 & 0.066319 & 53.1 &  FUV \\
JO194 & Abell 4059  & 359.2528333 & -34.680588 & 0.041951 & 38.7 &  FUV  \\
\hline
\end{tabular}  \\ 
\label{tab:Tab01}
\end{table*}

The FUV is observed for all of these galaxies, while NUV is observed only for JO201, JO60, and JW100.
Introduction and some analysis of the UV emission were previously done for JO201 (\citealt{Bellhouse17}, \citealt{George18}, \citealt{George19}), and JW100 (\citealt{Poggianti19b}).  
More recently, \citet{George23} presented the UV images of JO60,  JW39, and JO194 for the first time.  
These papers mostly compare the coverage of H$\alpha$ and UV tracers in detail, so this will not be a major topic and discussion in this paper (some minor comments on the subject in Sec. \ref{subsec:coverage}). 
The comparison in the emission values of H$\alpha$ and UV, and their corresponding SFR values, were done by \citet{George18} and \citet{George23}. 
However, we emphasize that these comparisons in the emission values were done for larger aperture areas (encompassing star-forming clumps and segments), with integrated spaxels, while this paper will concentrate on the pixel-by-pixel comparison at the highest resolution.

%%%%
%%   Subsec
\subsection{Optical IFU}\label{subsec:Data, MUSE}

The galaxies in this work were observed by optical IFU (MUSE\footnote{The Multi Unit Spectroscopic Explorer; \citet{Bacon10}.}) in order to trace their optical emission from the stellar continuum and the emission gas lines (nebular lines) from the ionized gas.  
The ionized gas is instantly ionized by massive, young ($<10$\,Myr) stars in star-forming regions (H$\textsc{ii}$), while the diffuse ionized gas  (DIG) is ionized partly by escaped ionizing photons from H$\textsc{ii}$ regions and partly by other sources of ionization (older stellar population, cosmic rays, shocks, and mixing of warm and cold gas layers, etc., \citealt{Haffner09}, \citealt{Tomicic17}, \citealt{Tomicic21b}).

The full description of the optical IFU observation and corresponding data analysis is described in detail by \citet{Poggianti17} and \citet{Fritz17}. 
We present here a short description of the procedures.
The IFU observations were done using MUSE at ESO-VLT \footnote{The Very Large Telescope of the European Southern Observatory.}, with spaxel sizes of $0.2''$ . 
The data calibration was done following the standard procedures. 
Due to the seeing effects during the observations, the calibrated data cubes were smoothed and convolved in the spatial dimension using a $5\times5$ pixel kernel, which corresponds to $1''$ (or $\approx 1$\,kpc at the galactic distances). 
The calibrated data were corrected for the foreground Milky Way extinction using the extinction values from \citet{Schlafly11}, assuming  \citet{Cardelli89} extinction curve and $\rm R_V = 3.1$.
The IFU cubes were then analysed by SINOPSIS (\citealt{Fritz17}) and  KUBEVIZ (\citealt{Fossati16})  to separate and fit the stellar and ionized gas emission components of the spectra.

Of the various emission lines derived from the spectra, we concentrate in this work on the two major Balmer lines, H$\alpha$ and H$\beta$ to derive Balmer line attenuation ($\rm A_V$), and SFR surface density, $\rm \Sigma(SFR)$.  
We assume the intrinsic Balmer line ratio $\rm H\alpha/H\beta=2.86$,  case B recombination and the gas temperature of $\rm T\approx10^4$\,K  as typically used in the literature (\citealt{HummerStorey87}, \citealt{Osterbrock92}).
Furthermore, we use other nebular lines to derive the Baldwin, Phillips \& Terlevich diagnostic diagrams (BPT, \citealt{Baldwin81}, \citealt{Kewley06})  diagrams that infer the dominating source of ionization in each spaxel. 
The stellar disks, their inclinations and centers were derived using optical contours that are 1\,$\sigma$ above the average sky background noise, and are described in detail by  \citet{Gullieuszik20} and \citet{Franchetto20}.
The tails of the galaxies are defined as areas outside the stellar disk.

In this paper, we only show and use pixels with signal-to-noise cut of  $\rm S/N\geq$4 for the Balmer lines in our analysis and for our results. 
We further separate pixels according to what fraction of the H$\alpha$ emission is emitted by the DIG, labeled with \cdig value that is defined by \citet{Tomicic21a}.

In the GASP survey, it is not possible to clearly separate \hii regions from the DIG due to constraints on spatial resolution. 
\citet{Tomicic21a} and \citet{Tomicic21b} estimated a \cdig  fraction, and we refer to regions dominated by \hii emission by the term ``dense gas'' in this paper.
\cdig value is measured by comparing [S\textsc{II}]/H$\alpha$ line ratio with H$\rm \alpha_{corr}$. 
The DIG-dominated regions exhibit higher [S\textsc{II}]/H$\alpha$ and low H$\rm \alpha_{corr}$, while the dense-gas dominated regions show lower [S\textsc{II}]/H$\alpha$ and high H$\rm \alpha_{corr}$.
These changes in the line ratios are predominantly due to an increase in the electron temperature and lower ionization parameter of the DIG  (\citealt{Haffner09}). 
Additional effects due to changes could not be excluded, such as changes in excitation sources (shocks and stellar feedback) and ionization conditions (the number and the population of ionizing photons).
Here we define the dense gas dominated pixels as the ones dominated by H$\textsc{ii}$ associations and with $\rm C_{DIG}<=0.5$, while the DIG dominated pixels as those with $\rm C_{DIG}>0.7$. 
We note that we are unable to fully separate H$\textsc{ii}$ from DIG emission due to low spatial resolutions (1\,kpc scales), but could use \cdig values as an approximate fraction of DIG emission in line of sight (LOS).

%%%%
%%   Subsec
\subsection{UV observations }\label{subsec:Data, UV}

The UV emission emerges directly from the photosphere of stars, which are mostly older than timescales of star-forming regions ($>10$\,Myr), with FUV tracing stars younger than 100\, Myr and NUV tracing stars up to  200\,Myr of age (\citealt{Kennicutt98b}).

The UV observations were performed  with the Indian multi-wavelength astronomy satellite (ASTROSAT)   and with ultra-violet imaging
telescope (UVIT) on it (\citealt{Agrawal06}, \citealt{Tandon17}, \citealt{Tandon20}). 
Details of the observations and data calibration is described in detail in \citet{George18} and \citet{George23}. 
The observations result in UV images with pixel scale of $\approx 0.4''$, and angular resolutions (the full-width of half maximum, FWHM) of $\approx1.4''$ and $\approx1.2''$ for FUV and NUV filter channels , respectively. 
The observations are photon-count based  with the poison statistics (\citealt{Tandon17}), and we measured the instrumental noise directly as a square root of the signal.
We emphasize that, the NUV observations have integration time twice as long as the FUV counterparts (for details see \citealt{George23}), which yields deeper NUV  observations with better S/N of the data compared to FUV. 
UVIT NUV channel stopped working in March, 2018 and we have only FUV data for JW29 and JO194 (\citealt{Ghosh21}).

The observed UV data have been convolved by the convolution kernel with the FWHM equal to the UV angular resolution, in order to estimate the background level across the pixels with no photon  counts (as similarly done by \citealt{George18a}).   
To measure the background level, we picked ten apertures, $11''$ in radius, outside the galaxies and their tails, and used their median value as a background value that was then subtracted from the image.
In the next step, we measured the noise level in fluxes as a standard deviation of the noise in those apertures outside the galaxies. 
The UV data were corrected for the  Milky Way dust extinction using the extinction values from \citet{Schlafly11}, assuming  \citet{Cardelli89} extinction curve and $\rm R_V = 3.1$.
We correct UVIT astrometry by comparing the position of bright stars in UV images with their position in the r-band images from the MUSE IFU cubes and the Hubble Space Telescope (HST) images. 
The uncertainty in the astrometry was 0.2$''$, which is equivalent to the MUSE pixel size.

%%%%
%%   Subsec
\subsection{Convolution and pixel transformation  }\label{subsec:Rebinning}

Before analyzing maps from different tracers, we first matched the pixel grid of maps from the optical IFU with that of the UV observations by pixel transformation. 
The matching of the images requires first the convolution of the optical maps due to different instrumental point-spread functions (PSF), and then the rebinning of pixels to match those of UV pixels.   
The proper pixel transformation is important in order to compare the coverage of the optical emission lines and UV continuum and to measure the proper ratios of those traces.

The MUSE images have PSF equivalent to the seeing (with FWHM$\approx 1''$). 
The instrumental FWHM of the UV data is 1.4$''$ for FUV and 1.2$''$ for NUV (\citealt{Tandon17}). 
The pixel scale of the MUSE and UVIT data are  0.2$''$ and 0.42$''$, respectively.

We first convolved the emission line maps of the galaxies using the \textit{numpy.convolve} PYTHON function and convolution kernel with $\sigma$ of the PSF equal to $\rm \sigma =\sqrt{\sigma_{UVIT}^2-\sigma_{MUSE}^2}$. 
Then we transformed the resulting convolved optical maps to match the UVIT maps by re-binning and changing the orientation of pixels, and interpolating their intensity values. 
While transforming the emission line maps, we estimated uncertainties in new maps by following the standard error propagation rules (adding uncertainties in quadrature).

To test if the process of pixel transformation (PSF convolution and re-binning) does not artificially change the H$\alpha$/UV flux ratios, we re-binned the H$\alpha$ and UV maps to pixel length of 1.4$''$ that is equal to the PSF size of the UV data.    
We confirm that there are no significant changes in H$\alpha/$UV ratios compared to the maps with smaller pixel sizes. 

Similarly, we derive the \cdig fraction maps by multiplying the original H$\alpha$ and \cdig maps at MUSE resolution,  and then convolving and re-binning the resulting H$\alpha_{DIG}$ maps at the UV resolution as described in Sec. \ref{subsec:Rebinning}.

\subsection{Conversion from tracers to SFR}\label{subsec:Data eta}

The SFR values are  estimated using three parameters: 1) the observed tracer $\rm f_{\lambda, obs}$, 
 2) the attenuation value $\rm A_\lambda$ to correct  the tracer for extinction effect, and 3) the SFR conversion factor $\rm \eta_\lambda$, as in the formula:

\begin{equation}
 \rm SFR_{f_{\lambda}} = \eta_\lambda \times f_{\lambda, obs} \times 10^{0.4A_\lambda}. 
\label{eq:Eq_ana_1}
\end{equation}

We  adopt the  $\eta$ conversion factors for the UV and H$\alpha$ tracers (labeled with $\rm \eta_{UV}$ and $\rm \eta_{H\alpha}$) to covert  surface brightness of the tracers to SFR surface densities $\rm \Sigma(SFR)$  as:
\begin{equation}
\frac{\rm \Sigma_{SFR}(Tracer)}{\rm M_{\odot}yr^{-1}kpc^{-2} }=\eta\,\cdot\frac{\rm \Sigma(\rm Tracer)_{corr}}{\rm erg\,s^{-1}kpc^{-2}}
\label{eq:eta general}
\end{equation}

The values of $\Sigma(\rm Tracer)_{corr}$ are attenuation-corrected surface brightness of the tracers (\citealt{Calzetti94}). 
We use the measured H$\alpha$/H$\beta$ line ratios and the \cite{Cardelli89} extinction curve to estimate the H$\alpha$ attenuation value ($\rm A_{H\alpha}$). 
For the UV tracers, we use $0.44\times$ the value of the \cite{Cardelli89} extinction curve because the UV stellar attenuation curve is grayer (i.e. lower) than the nebular extinction curve (\citealt{Calzetti94}, \citealt{Calzetti00},  \citealt{Calzetti07}, \citealt{Kreckel13}).
This yields UV attenuation values  of $\rm A_{UV}\approx 1.38\times A_{H\alpha}$. 

The conversion factors $\eta$ for different tracers are taken from \citet{KennicuttEvans12} (see their Table 1), with values of $\eta(H\alpha) = 5.37\times 10^{-42}$, $\eta(FUV) = 4.47\times 10^{-44}$, and $\eta(NUV) = 6.76\times 10^{-44}$. 
These values were estimated by \citet{Murphy11} and \citet{Hao11} where they used the Starburst99\footnote{http://www.stsci.edu/science/starburst99/docs/default.htm} stellar population models. 
These conversions also assume a constant star formation over 100 Myr, and the Kroupa initial mass function (IMF; \citealt{Kroupa01}) of formed stellar clusters.

\begin{figure*}
\centering
\includegraphics[width=1\textwidth]{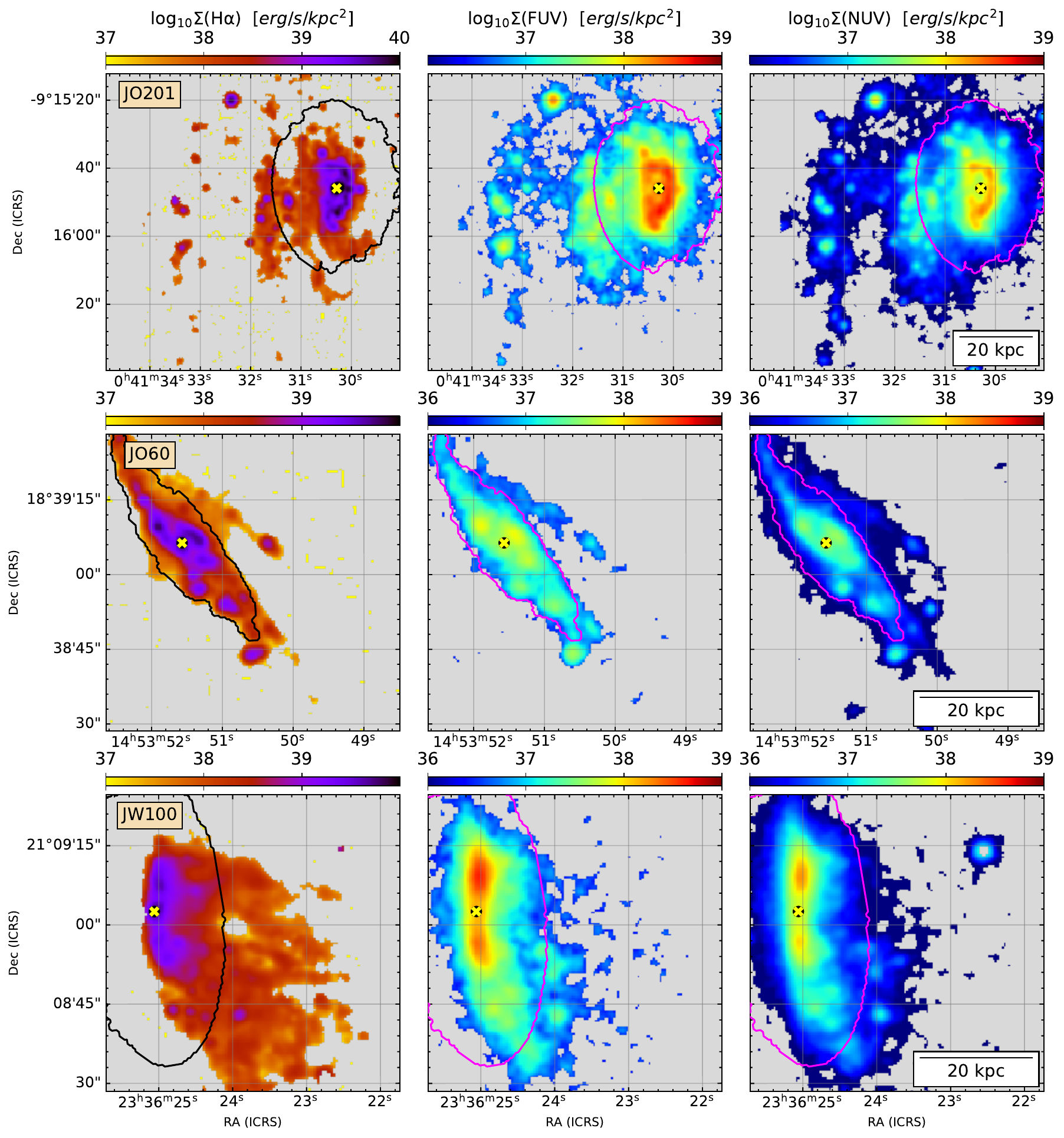}
\caption{Observed surface brightness maps of H$\alpha$ emission line (left column), FUV (middle column), and NUV (right column) of galaxies JO201, JO60, and JW100 (from top to bottom respectively). The stellar disks are indicated by thick contours, and their disk centers are marked with yellow crosses on the black circle. Here, the H$\alpha$ tracer maps are matched with the UV angular resolution and pixel sizes. The S/N threshold used for this figures is $\rm S/N>=4$ for the H$\alpha$ and the UV maps.}    
\label{fig:AllGal_mapHaUV}
\end{figure*}

\begin{figure}
\centering
\includegraphics[width=0.5\textwidth]{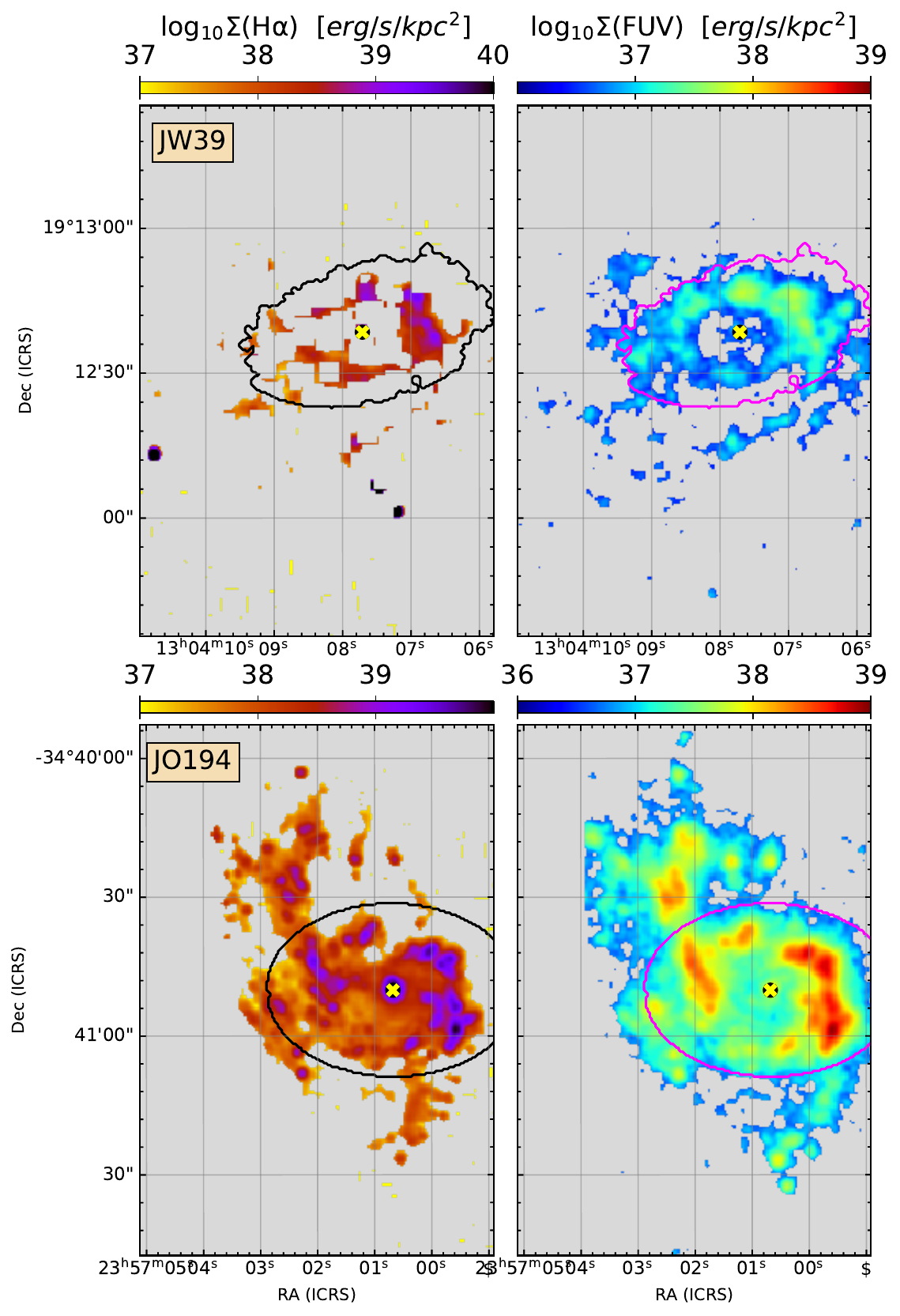}
\caption{Same as Fig. \ref{fig:AllGal_mapHaUV} for JW39 and JO194. In the case of these galaxies, we observed only FUV emission.  }
     \label{fig:AllGal_mapHaUV_p2}
\end{figure}

\begin{figure*}
\centering
\includegraphics[width=1.\textwidth]{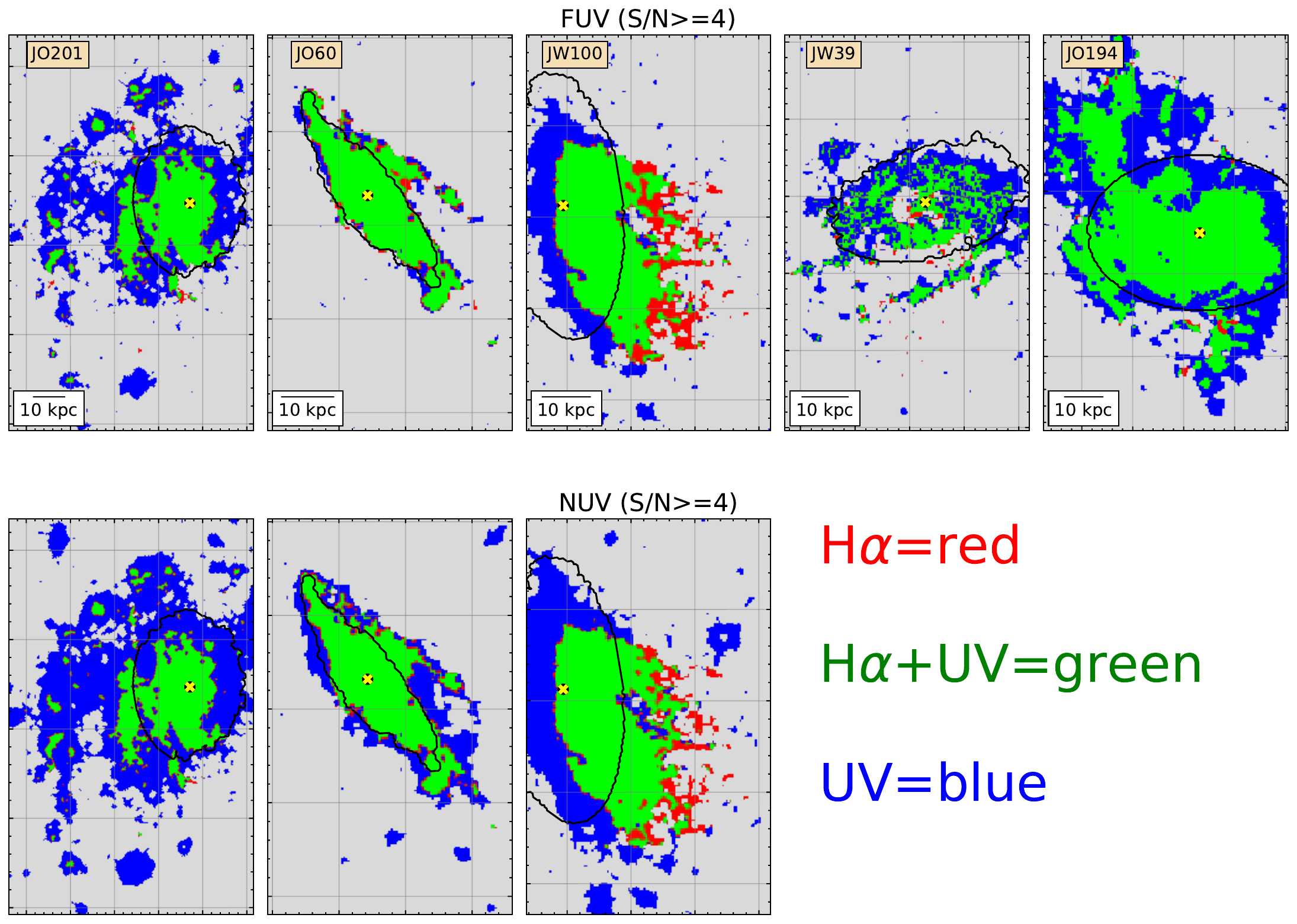}
\caption{Coverage of the optical and UV tracers (FUV on top, and NUV on bottom) for the galaxies.  The S/N threshold used for the tracers is $\rm S/N>=4$.}  
 \label{fig:AllGal_mapHaUV_cov}
\end{figure*}

\begin{figure*} 
\centering
\includegraphics[width=0.95\textwidth]{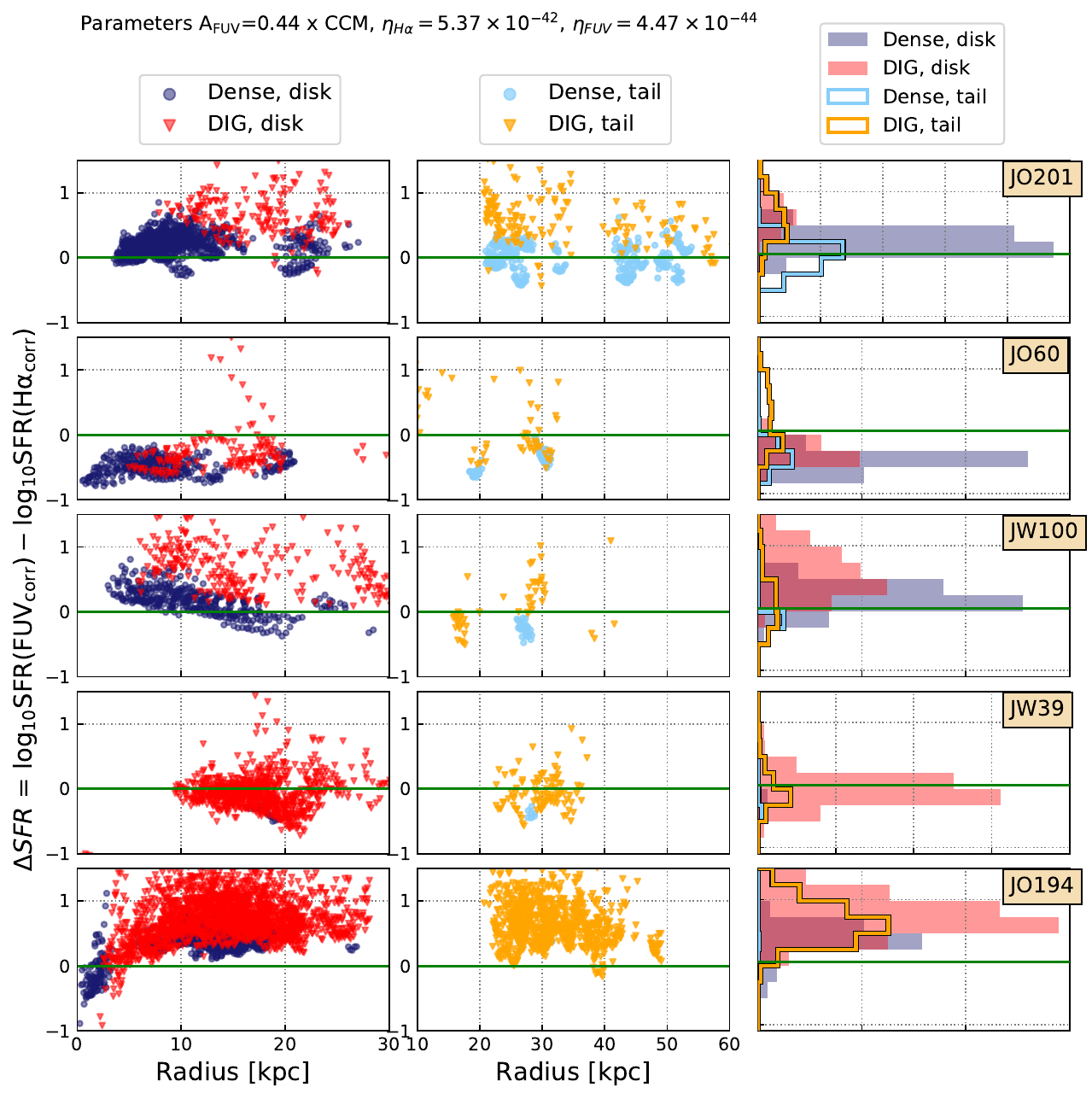}
\caption{ $\rm \Delta SFR = log_{10}SFR(FUV) - log_{10}SFR(H\alpha)$  for pixel-by-pixel values (y-axis) as a function of galactocentric distance (left and center panels, x-axis). We present data for the disks (tails) in the left (center) panel, respectively. The histogram of the $\rm \Delta SFR$ (normalized units) is on the right panel.   We define the dense gas-dominated pixels with $C_{DIG}<=0.5$ (dark blue for disks, and light blue for tails) and the DIG-dominated pixels with $C_{DIG}>=0.7$ (red for disks and orange for tails). The UV attenuation values are estimated as $0.44$ of the extinction values using the \citet{Cardelli89} extinction curve (CCM).   }
     \label{fig:diffSFR rad hist FUV}
\end{figure*}

\begin{figure*}
\centering
\includegraphics[width=0.8\textwidth]{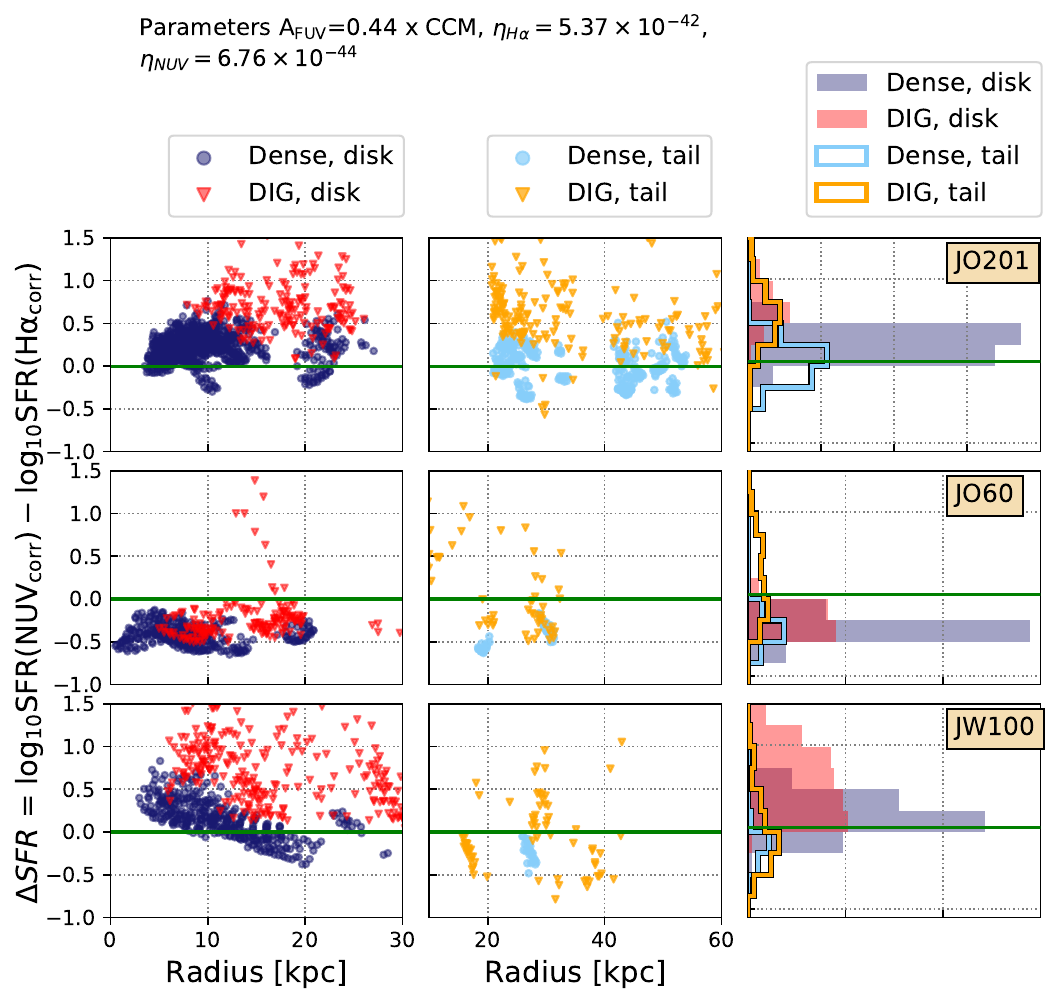}
\caption{ Same as Fig. \ref{fig:diffSFR rad hist FUV} but for NUV tracer.  }
     \label{fig:diffSFR rad hist NUV}
\end{figure*}

%%%%%%%%%%%%%%%%%%%%%%%%%%%
%%%%
%%   Results
%%%%
%%%%%%%%%%%%%%%%%%%%%%%%%%
\section{Results} \label{sec:Results}

%%%%
%%   Subsec
\subsection{Observed emission maps}\label{subsec:UV-Ha maps}

We show the H$\alpha$, FUV and NUV images of the 5 galaxies in Fig. \ref{fig:AllGal_mapHaUV} and \ref{fig:AllGal_mapHaUV_p2}. 
In the rest of the paper, we will compare pixel-by-pixel SFR values from the H$\alpha$, FUV, and NUV tracers. 
We only consider those pixels that show both the H$\alpha$ emission and UV tracers with $\rm S/N>=4$ and are powered by SF according to the BPT-$\rm [OI]$ diagram. This BPT diagram uses the line ratio of  $\rm [O\textsc{I}\lambda 6300]/H\alpha$ (\citealt{Kewley06}). 
Furthermore, we compare disks and tails of stripped galaxies, dense gas and DIG-dominated pixels.

Based on visual inspection, the UV and H$\alpha$ are mostly spatially coincident.  
We note that the NUV tracer covers a larger area than the FUV tracer.
This is probably due to a shorter observation time conducted for  FUV compared to NUV, thus FUV has lower S/N compared to NUV. 
Using the uncertainties of the maps (not corrected for attenuation), we estimated approximate lower limits in $\Sigma$SFRs that maps would show: $\approx 5\times 10^{-5}\, M_\odot /kpc^{2}$ for FUV, $\approx 2.2\times 10^{-5}\, M_\odot /kpc^{2}$ for NUV, and $\approx 3.7\times 10^{-5}\, M_\odot /kpc^{2}$ for H$\alpha$.  
In Fig. \ref{fig:AllGal_mapHaUV_cov}, we show the spatial coverage of different tracers in the galaxies, showing that most of the areas with H$\alpha$ is also covered with the UV tracers. 
As expected from the estimated lower SFR limits of the tracers, NUV covers a larger area compared to H$\alpha$ and FUV.  
We further discuss differences in the coverage areas of the tracers in Sec. \ref{subsec:coverage}.

%%%%
%%   Subsec

\subsection{Comparison  between SFR(UV) and SFR(H$\alpha$)}\label{subsec:diff SFR}

In Fig. \ref{fig:diffSFR rad hist FUV} and \ref{fig:diffSFR rad hist NUV} we show  a pixel-by-pixel comparison of $\rm \Delta SFR$ for the galaxies. 
For each galaxy, we separate data according to the DIG emission fraction and separate disks (darker colors) and tails (light colors).   
In the left columns of Fig. \ref{fig:diffSFR rad hist FUV} and \ref{fig:diffSFR rad hist NUV}, we compare the difference in SFRs as a function of galactic radius, and in the right columns the distribution of the data (filled histograms for disks, and empty for tails).
With the investigation of radial behavior of  $\rm \Delta SFR$, we aim to probe the effects of environments, such as central and outer disks and tails at different distances from the stripping effect.

Focusing on the dense gas first, we note that dense-gas spaxels have a range of $\rm \Delta SFR$ within 0.5 dex around the value 0. 
This observation is similar to the one made by \citet{George18} and \citet{George23}, where they observe $\Delta SFR\approx 0$ for the integrated values of star-forming clumps in the galaxies.  
This is expected due to the definition of the dense gas spaxels tracing the regions of recent star formation.  
However, a large galaxy by galaxy variation exists, where JO60 and JW39  show $\rm \Delta SFR<0$. 
We notice a small variation (0.5\,dex drop) with the galactocentric radius in JW100,  and no difference between the disks and tails.
The center of JO194 shows a stark drop in $\rm \Delta SFR$.

Considering  DIG-dominated pixels, we find higher values in SFR(UV) than in SFR(H$\alpha$) ($\rm \Delta SFR\approx$0.5 dex higher), with a large scatter.
We note that JW100 shows some decline in $\rm \Delta SFR$ as a function of the galactocentric radius, where the tails have $\rm \Delta SFR\approx 0$. 
Furthermore, JO60 also shows $\rm \Delta SFR\approx 0$ for the DIG-dominated data. 

There is a clear distinction in $\rm \Delta SFR$ between the dense gas and DIG-dominated spaxels ($0.5-1$\,dex), which is larger than variations between the galaxies or between the disks and tails. 
The DIG-dominated spaxels show similar values for the disks and tails. 

\begin{figure*}
\centering
\includegraphics[width=0.9\textwidth]{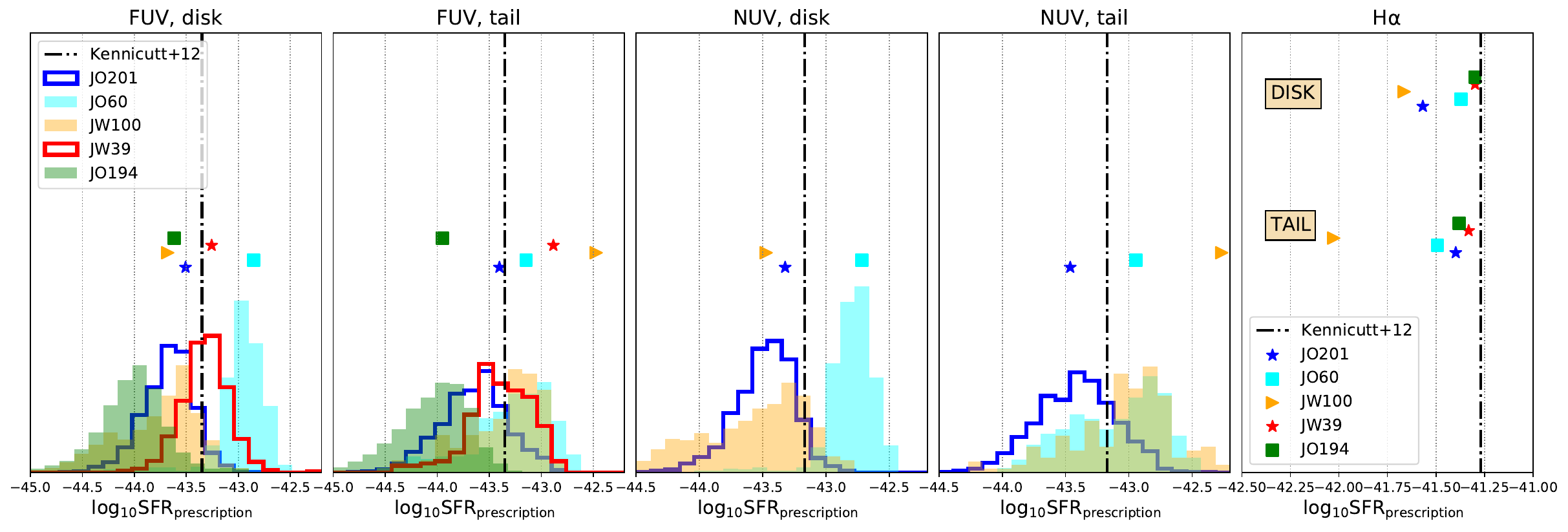}
\caption{ Empirically estimated SFR prescriptions $\eta$ for FUV (1st and 2nd panels), NUV (3rd and 4th panels) and H$\alpha$ (5th panel),  for the disks and the tails. We show pixel-by-pixel comparison as histograms (except the case for H$\alpha$), and integrated values as single data points. We also indicate the prescriptions from the literature  (\citealt{KennicuttEvans12}) by the vertical, dashed line.  }
     \label{fig:AllGal_SFR prescriptions}
\end{figure*}

%%%%
%%   Subsec
\subsection{Empirical SFR prescriptions for UV}\label{subsec:amp SFR prescriptions}

We now empirically estimate the SFR prescriptions for disks and tails of stripped galaxies, by using SFR(H$\alpha_{corr}$)  as reference values.

To do that, we use the relation:
\begin{equation}
 \rm log_{10}SFR(H\alpha_{corr}) = C_\lambda + log_{10}L_\lambda, 
\label{eq:Eq prescr}
\end{equation}

where $\rm C_\lambda$ is the empirical prescription, and $\rm SFR(H\alpha_{corr})$ and $\rm L_\lambda$ are measured SFRs and luminosity of the extinction corrected tracers (FUV, NUV and $\rm H\alpha$). 
For the measured $\rm SFR(H\alpha_{corr})$, we only consider spaxels defined by BPT-[O\textsc{I}] as star-forming, while for $\rm L(H\alpha)$  we also consider non-star-forming spaxels. 
This will result in $\rm SFR(H\alpha_{corr})$ corresponding only to more luminous spaxels (with sufficiently bright H$\beta$), while $\rm L(H\alpha)$ spaxels also including low-intensity emission from highly attenuated regions, low S/N(H$\beta$) spaxels, or gas not ionized by star formation.

\begin{figure*}
\centering
\includegraphics[width=0.9\textwidth]{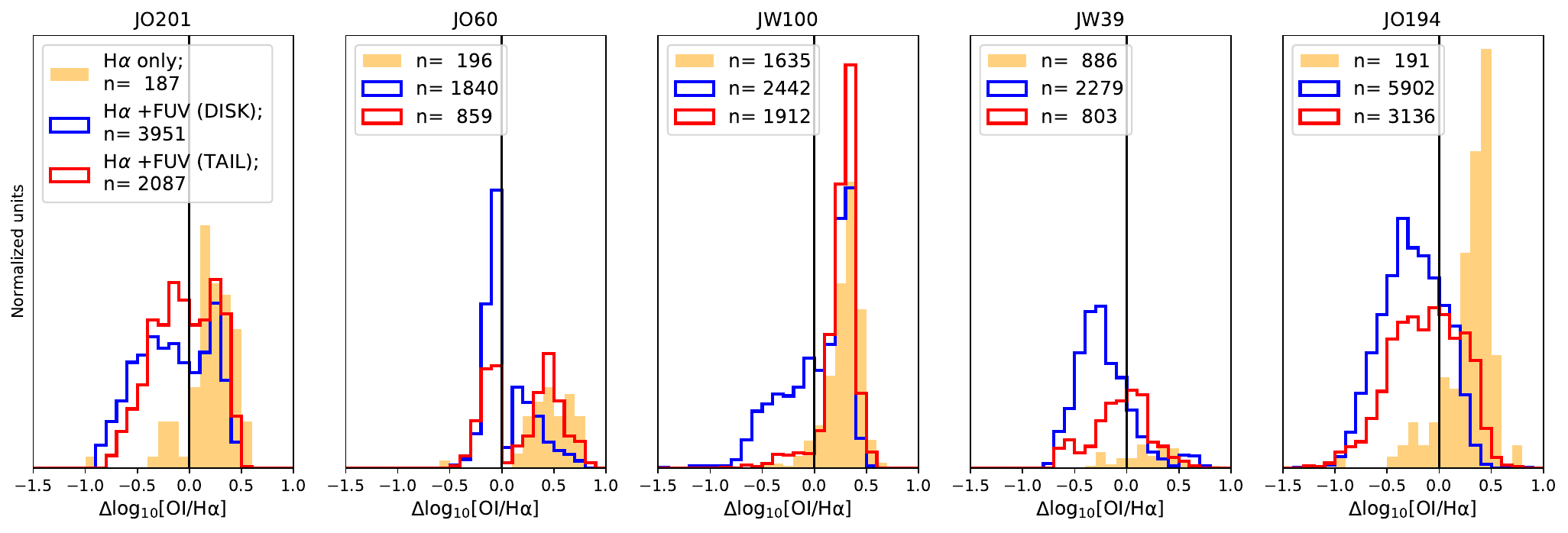}
\caption{Distribution of the number of spaxels characterized by a different \doha in the galaxy regions, characterized by H$\alpha$ only, and H$\alpha+$FUV.  The number of spaxels is on the y-axis, normalized by pixel numbers.  H$\alpha$-only areas are presented with filled orange histograms, H$\alpha+$FUV areas in the disks (tails) with empty histograms with blue (red) edges, respectively. \doha value is defined  as an offset in $\rm log_{10}[O\textsc{I}]/H\alpha$ from the star-forming line in the BPT-[O\textsc{I}] diagram by \citet{Tomicic21b}. Positive and negative \doha values indicate that the gas was predominantly ionized by the SF and non-SF processes, respectively.  In the panels, we also list the number of pixels in those areas. The black, vertical line indicates value $\rm log_{10}[O\textsc{I}]/H\alpha=0$ and separates SF from non-SF spaxels.   }     \label{fig:AllGal_histo_Coverage_doha}
\end{figure*}

The measurement of $\rm C_\lambda$ (Fig. \ref{fig:AllGal_SFR prescriptions} and Tab. \ref{tab:Tab presc}) was done on a pixel-by-pixel basis (histograms) and on integrated values (data points),  within the disks and tails separately.   
We also show the prescription values from the literature (\citealt{KennicuttEvans12}).  
We note that the pixel-by-pixel base data show a large scatter and range in $\rm C_{UV}$ values, with galaxies differing up to 1 dex between each other.   
In most cases,  disks and tails of the same galaxies have a similar range in $\rm C_{UV}$. 
The integrated values mostly have similar  $\rm C_{UV}$ compared to the peaks of pixel-by-pixel distributions.
$\rm C_{UV}$ values for different galaxies mostly exhibit lower values compared to the literature, while the literature prescriptions are still within the data scatter from the mean of $\rm C_{UV}$  (Tab. \ref{tab:Tab presc}). 
Integrated data for $\rm C_{H\alpha}$ shows 0.25\,dex higher values for disks compared to the tails of most galaxies, except for JW100. 
This is due to a large fraction of H$\alpha$ emission in its tail being dominated by  DIG that is not ionized by star-forming photons, thus shifting the prescription to lower values. 
The mean value of $\rm C_{H\alpha}$ is lower by $\approx 0.2$\,dex compared to the literature value.

 \begin{table}[h!]
\centering
\caption{ The empirical SFR prescriptions  C$_\lambda$ for FUV, NUV and H$\alpha$ from the spatially resolved (middle column)  and integrated data (right column), for disks and tails. The SFR prescriptions are defined as   $\rm log_{10}SFR(H\alpha_{corr}) = C_\lambda + log_{10}L_\lambda$. \citet{KennicuttEvans12} uses C$_\lambda$  equivalent to  $-43.35$,  $-43.17$ , and $-41.27$  for FUV, NUV and $\rm H\alpha$ respectively. Added uncertainty is scatter of the pixel-by-pixel data, and the variance of galaxies for the integrated data. There are no values for the spatially resolved C$_{\rm H\alpha}$ because in that case H$\alpha$ and H$\rm \alpha_{corr}$ spaxels are the same spaxels.  }
\begin{tabular}{l|l|l} 
\hline \hline
 &  Spatially resolved C$_\lambda$ & Integrated C$_\lambda$ \\
\hline 
FUV (disk) & $ -43.5\pm0.3 $   &  $ -43.38\pm0.09 $      \\
FUV (tail) & $ -43.4\pm0.4 $   &  $ -43.1\pm0.2 $  \\
NUV (disk) & $ -43.2\pm0.3 $   &  $ -43.2\pm0.1 $  \\
NUV (tail) & $ -43.1\pm0.4 $   &  $ -42.9\pm0.2 $  \\
$\rm H\alpha$ (disk) & &  $ -41.44\pm 0.02 $   \\
$\rm H\alpha$ (tail) & &  $ -41.52\pm 0.07 $  \\
\hline
\end{tabular}  \\ 
\label{tab:Tab presc}
\end{table}

%%%%%%%%%%%%%%%%%%%%%%%%%%%
%%%%
%%   Discussion
%%%%
%%%%%%%%%%%%%%%%%%%%%%%%%%
\section{Discussion} \label{sec:Discussion}

Our main results (Fig. \ref{fig:diffSFR rad hist FUV} \&  \ref{fig:diffSFR rad hist NUV}) indicate that SFR prescriptions for UV and H$\alpha$ in the disks and tails of the stripped galaxies do not differ, but they do differ between the dense gas and DIG dominated areas. 
This is in agreement with findings of \citet{George18} and \citet{George23}, where they observe the same SFR prescriptions for the integrated star-forming knots in both the disks and tails of the GASP galaxies. 
The SFR(UV)  dominates over  SFR(H$\alpha$) in the DIG-dominated regions across the galaxies, and its potential causes are further discussed in the following subsections.

%%%%%
%% Subsec
\subsection{Non-SF areas}\label{subsec:coverage}

Some discrepancies between SFR may arise from the non-SF nature of ionization in H$\alpha$ and due to different timescales of SF that tracers probe, about which we further discuss in Sec. \ref{subsec:SF time}.
In Fig. \ref{fig:AllGal_mapHaUV_cov}, we showed that galaxies exhibit some areas where there is only UV emission and no H$\alpha$ (blue in the maps). 
Similar UV-only features in our sample have been observed with imaging by the Hubble Space Telescope by \citet{Giunchi23}.  
In particular, JW100 and JW39 (and partly other galaxies) show UV-only areas at the ram-pressure fronts where they are colliding with the ICM gas (\citealt{Bellhouse17, Poggianti19b}), indicating areas where the gas might potentially be stripped out from the stellar disk and transferred to the tails.
The leading edge of jellyfish galaxies tend to show post-starburst regions in their spectra, as shown by \citet{Gullieuszik17}, \citet{Poggianti19b}, \citet{Werle22} and \citet{Werle23}.
The UV and  H$\alpha$ tracers may spatially differ across the galaxies if there is a spatial separation between stars of different ages.

 We note that all galaxies exhibit some more diffuse areas in the tails with H$\alpha$ emission only (red in the maps), and this feature is most noticeable in JW100.
This can be due to the spatial separation of very young and older stars, but that could also be due to the DIG in those regions being ionized by processes other than SF (\citealt{Tomicic21a, Tomicic21b}). 
If the H$\alpha$ emission-only regions have the gas ionised by such processes, they would yield higher line ratios of $\rm [O\textsc{I}\lambda6300]/H\alpha$ (\citealt{Kewley06}), and even the $\rm [O\textsc{II}\lambda3727]/H\alpha$ excess in the stripped tails (\citealt{Moretti22}).

To test this, we show histograms of \doha values for H$\alpha$-only and H$\alpha+$FUV (for the disks and tails) areas in Fig. \ref{fig:AllGal_histo_Coverage_doha}.
\citet{Tomicic21a} and \citet{Tomicic21b} defined a \doha value as an offset in $\rm log_{10}[O\textsc{I}]/H\alpha$ from the star-forming line in the BPT-[O\textsc{I}] diagram, to investigate if stripped galaxies show statistically higher $\rm log_{10}[O\textsc{I}]/H\alpha$ fractions compared to non-stripped galaxies.  
Positive \doha values indicate LINER emission.
For all the galaxies, H$\alpha$-only areas exhibit higher \doha values compared to H$\alpha+$UV areas and mostly show positive values indicating that the gas in those areas is indeed most likely ionized by processes other than SF. 
We also note that  H$\alpha+$FUV for the disks and tails cover a similar range in \doha values, around 0, with the disk distribution having more negative values compared to the tail values.
The only outlier is the tail of JW100, whose \doha distribution of the H$\alpha+$UV region has mostly positive values.
 
These results indicate that a process other than SF  can affect the SFR values measured in the tails of the stripped galaxies. 
One such non-SF process could be the mixing of the hot ICM and cold ISM in tails, where the thermal, photoionizing radiation of the ICM can stimulate H$\alpha$ emission in the stripped ISM. This idea was brought previously by various papers regarding the ISM medium in stripped galaxies (\citealt{Slavin93, Binette09, Poggianti19b,Sparre20, Campitiello21, Muller21, Franchetto21, Bartolini22, Sun22, Khoram24}).

%%%%
%%   Subsec
\subsection{How does attenuation affect SFR discrepancies} \label{subsec:var SFR p2}

The assumed attenuation curve may affect the difference in SFR estimates between UV and H$\alpha$, as different curves will produce very different effects in the UV while keeping H$\alpha$ more or less unaffected (at constant $\rm A_V$). 
Furthermore, the absolute value of $\rm A_V$ also partly affects the $\eta$ parameter.
Therefore, here we test if the assumption of the attenuation curve affects SFR values as significantly as uncertainty in the $\eta$ parameters.

Applying Eq. \ref{eq:Eq_ana_1} to the difference in SFR between UV and H$\alpha$ tracers, we  derive the difference as:

\begin{equation}
\begin{split}
  \Delta SFR = \text{log}_{10}(\frac{\lambda_{FUV}f_{FUV, obs}}{f_{H\alpha, obs}}) + \\  
  \text{log}_{10}(\frac{\eta_{FUV}}{\eta_{H\alpha}}) + 0.4 \times (A_{FUV}-A_{H\alpha}) .
\end{split}
\label{eq:Eq_ana_2}
\end{equation}

The first term on the right side presents observable values and cannot be changed by our assumed SFR prescription. 
The second and third terms depend on our SFR prescription and assumptions of attenuation curves, thus affecting $\rm \Delta SFR$ values. 

 \begin{table}[t!]
\centering
\caption{Range of the parameters used in the analytical test in Sec. \ref{subsec:var SFR p2}. }
\begin{tabular}{ll} \hline \hline
Parameter & Range of the parameter \\
\hline 
$\eta_{FUV}$  & $(5 \pm 3)\times 10^{-44}$ \\
$\eta_{H\alpha}$  & $(5 \pm 2)\times 10^{-42}$ \\
$\rm 0.4\times (A_{FUV}- A_{H\alpha})$  & $\rm (0.47 \pm 0.28)\times A_V $ \\
\hline
\end{tabular}  \\ 
\label{tab:Tab02}
\end{table}    	

\begin{figure*}
\centering
\includegraphics[width=0.9\textwidth]{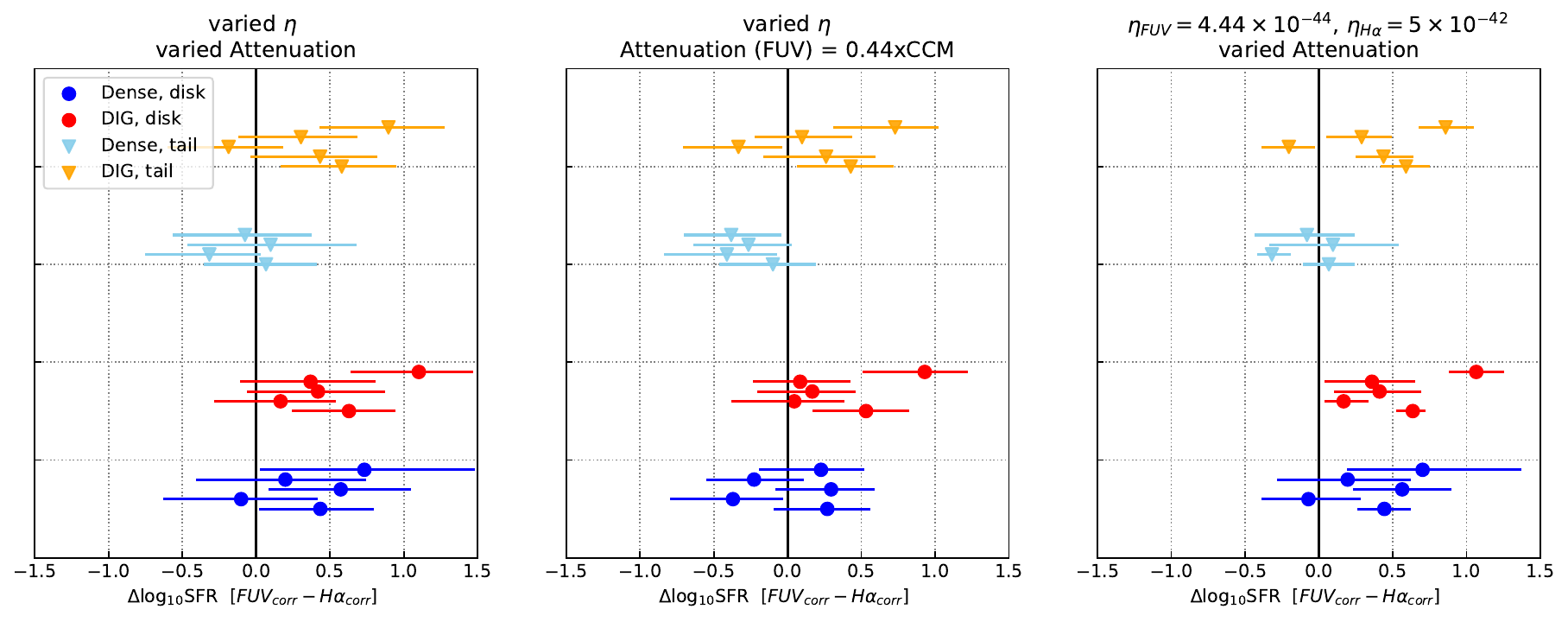}
\caption{  The Monte Carlo simulation of a scatter in $\rm \Delta SFR$ values due to variation in the SFR prescriptions and assumption of attenuation curve for UV tracer (see Eq. \ref{eq:Eq_ana_2}).    The presented cases are where the following parameters are varying: 1) both  $\eta$ and A$\rm _{UV}$ (left panel), 2)  only $\eta$ parameters and the $\rm 0.44x$\,CCM curve (\citealt{Cardelli89}) for UV is assumed (middle panel), and 3) the $\eta$ parameters are set and the attenuation curve for UV varies between $\rm 0.44x$\,CCM and $\rm 1x$\,CCM (right panel).   We show median value with the 25th and 85th percentile (horizontal lines) for the disk (circles) and tails (triangles), dense gas (blue colors) and DIG (red and orange colors) dominated spaxels in JO201, JO60, JW100, JW39, and JO194 (from top to bottom of  point clusters).       }
     \label{fig:Analitical variations}
\end{figure*}

We show a simple test (Fig. \ref{fig:Analitical variations}) of changes in $\rm \Delta SFR$ by taking observed values of UV and H$\alpha$, and varying $\eta$ and A$\rm _{UV}$ (attenuation curve for UV) parameters.  
We use the Monte Carlo method (MCM; \citealt{Metropolis_ULam49}) where we vary different terms in three cases: 1) we vary both  $\eta$ and A$\rm _{UV}$ (left panel), 2)   we vary only $\eta$ parameters and assume $\rm 0.44x A_\lambda/A_V$, where $\rm A_V$ is given by the Milky Way extinction curve of \citet{Cardelli89} (labeled as CCM) for UV (middle panel), and 3)  we set $\eta$ parameters and vary attenuation curve for UV between $\rm 0.44\times A_\lambda/A_V$ and $\rm 1\times A_\lambda/A_V$ (right panel).
For the CCM curve, we are using a value of $\rm R_V=3.1$ (the average value for the Milky Way), and the choice of 0.44 corresponds to the typical difference between old and young stellar populations in galaxy disks (\citealt{CharlotFall00}).  
The case of $\rm 0.44\times A_\lambda/A_V$ (labeled in Fig. \ref{fig:Analitical variations} as $\rm 0.44\times CCM$) represents examples where the attenuation of light is lower than in the case of a pure CCM extinction curve ($\rm 1\times A_\lambda/A_V$, labeled as $\rm 1\times CCM$), and it is caused by the case of dust and young stars being more mixed  (\citealt{Cardelli89, Calzetti00, Galliano18}).

The parameters are varied (Tab. \ref{tab:Tab02}) using a random sample from a normal distribution, where mean and $\pm \sigma$ values are:  $\rm \eta_{FUV}= (5 \pm 3)\times 10^{-44}$,  $\rm \eta_{H\alpha}= (5 \pm 2)\times 10^{-42}$ for the tracers. The approximate mean values of $\eta$ parameters were taken from Table 1 in \citet{KennicuttEvans12}, and the approximate ranges by using the Starburst99 model by \citet{Leitherer99}\footnote{ see details in \url{https://ned.ipac.caltech.edu/level5/Sept12/Calzetti/Calzetti1_2.html}}. For the range in attenuation values,  we used $\rm 0.4\times (A_{FUV}- A_{H\alpha}) = (0.47 \pm 0.28)\times A_V $ corresponding to a range between $\rm 0.44x$\,CCM and $\rm 1x$\,CCM. In the case of the CCM curve, attenuation values for the tracers would be  $\rm A_{FUV} = 2.6907\times A_V$ and $\rm A_{H\alpha} = 0.8178 \times A_V$. 

In Fig. \ref{fig:Analitical variations}, we show the median and the 15th and 85th percentiles (horizontal lines) of $\rm \Delta SFR$ (measured as in Eq. \ref{eq:Eq_ana_2}) that is a result of joining the observed spaxels with varying $\eta$ and attenuation parameters from the Monte Carlo simulation (where we run 1000 iterations for each case). 
The data show disks and tails of the galaxies, and dense gas and DIG-dominated spaxels, separately. 
The results show a wide range in $\rm \Delta SFR$ (up to 1 dex between the 15th and 85th percentile), which is caused by significant variations in $\eta$ and $\rm A_{UV}$ parameters (left panel). 
Nonetheless,  a larger scatter in $\rm \Delta SFR$ is observed in the case of varying $\eta$ only (0.5 dex variation; middle panel) compared to the case where we vary only attenuation assumption (variation mostly $<0.2$dex; right panel).
Note that varying attenuation curve assumptions for H$\alpha$ do not change results significantly.

These results indicate that this assumption in the SFR prescriptions (variation in $\eta$ parameters) leads to larger variations in measured $\rm \Delta SFR$ compared to the assumption of a different attenuation curve (CCM extinction curve vs. $\rm 0.44x$\,CCM that is similar to the UV attenuation law observed \citealt{Calzetti00}\footnote{\citet{Calzetti00}  measured for the star-burst galaxies values of $\rm A_{1600\AA}  \approx 4.39 \times E_{B-V} \approx 1.4 \times A_V$  (Eq. 10 in their work).}) for the UV tracer. 
Therefore, we conclude that our assumption of the attenuation curve for UV is relatively good and that the major driver in the scatter and deviation in $\rm \Delta SFR$ are driven by assumptions in SFR prescriptions ($\rm \eta_{UV}$ and $\rm \eta_{H\alpha}$), such as timescale of SF, IMF, escape fractions, etc.

\begin{figure*}
\centering
\includegraphics[width=1.\textwidth]{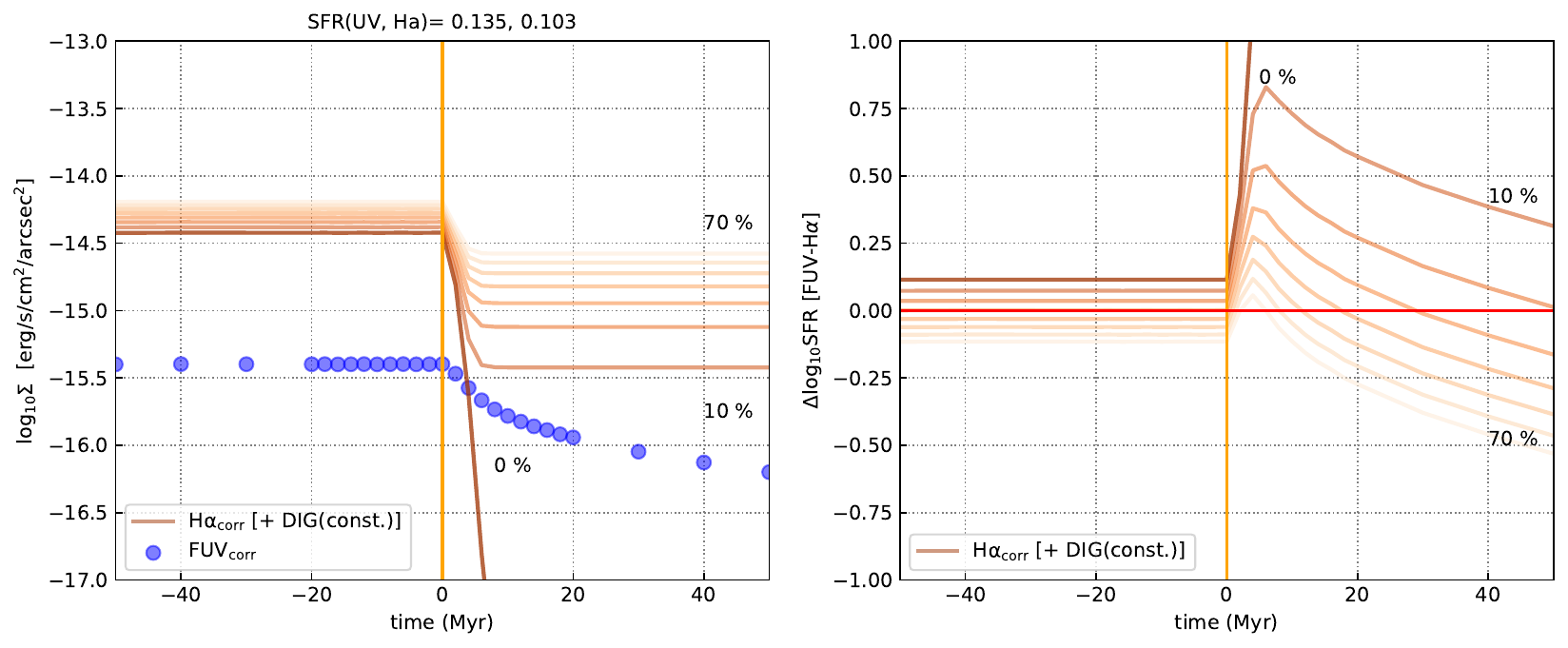}
\caption{ The model of star formation and UV (blue dots) and H$\alpha$ (red-orange-yellow lines) emission that it produces. The diagrams show the observed tracers (left panel) and the measured $\rm \Delta SFR$ (right panel) as a function of time. The SFR of the model is constant at t$<$0\,Myr, and stops at t$=$0\,Myr. We also added a separate  H$\alpha$ emission of DIG (as a percentage of the H$\alpha$ emission from the SF process), which is not affected by the drop in SFR, to mimic the existence of DIG that might not be fully ionized by SF process. We show a different intensity of DIG, from 0\% up to 70\% of H$\alpha$ from the SF process, as H$\alpha$ lines with a different shade of red color in the figure. 
 }
     \label{fig:Simulation}
\end{figure*}

%%%%
%%   Subsec
\subsection{Star-formation history}\label{subsec:SF time}

As mentioned before, UV and H$\alpha$ trace star-formation in different timescales, which is why variations in the recent star-formation history may play a role in shaping the difference in the SFRs obtained from these two tracers (\citealt{Cleland21, Ignesti22d}, \citealt{ArangoToro23, Edler23}).
To test this effect, we generated a star-formation history model using the {\sc bagpipes} code \citep{BAGPIPES} and followed the resulting evolution of FUV and H$\alpha$ emission.
The simulated star-formation history has a constant SFR for 500 Myr, after which the star-formation is abruptly quenched.
The resulting observables are illustrated in Fig. \ref{fig:Simulation}, where blue dots indicate UV and red lines show H$\alpha$ emission, and the time when star-formation is quenched is shown at t=0.
The model is based on the 2016 update to the \cite{Bruzual2003} stellar population models, with emission lines generated by Cloudy with ionization parameter $\log U=-2.5$, metallicity following the stellar value, and assuming ionization bound nebulae. 

We also added a separate  H$\alpha$ component (as a percentage of the H$\alpha$ emission from the SF process) that is not affected by the drop in SFR, to mimic the presence of DIG that might not be fully ionized by SF process.  
We show a different intensity of DIG, from 0\% up to 70\% of H$\alpha$ from the SF process, as H$\alpha$ lines with a different shade of red color in the figure. 
In the right panel of  Fig. \ref{fig:Simulation}, we show how changes in the tracers would affect the measurement of $\rm \Delta SFR$. 
The simulation indicates that the $\rm \Delta SFR$ increases toward positive values within the first 20-40 Myr after the end of star formation. 
This is due to a faster drop in H$\alpha$ after the first drop in the star-forming process and might be prevalent in regions outside the regions of recent star-formation.

Our new result is in line with the previous findings based on the study of radio continuum emission as a proxy for the SFR decline (\citealt{Ignesti22, Ignesti22c}), thus confirming that fast variations in the physical properties of galaxies, either induced by environmental or internal processes, can offset the different emission mechanisms we commonly use to trace the star formation (\citealt{Edler23, Roberts23}). 
The consequences are twofold. On the one hand, it may imply that the empirical SFR relation cannot be safely applied to these galaxies because they do not account for these processes. 
On the other hand, it raises the interesting prospect of combining multiple tracers, which probe the star formation rare with different timescales, to infer the time scale of SFR quenching, and hence, of the galaxy evolution.

From our simulations we conclude that the observed increase in $\rm \Delta SFR$ for the DIG-dominated regions is potentially caused by a recent (up to 40\,Myr ago) termination of star-formation. 
Those regions still exhibit both the diminishing UV emission from stars and H$\alpha$ emission from an additional DIG.

%%%%
%%   Subsec
\subsection{Other sources of the discrepancies }\label{subsec:other disc}

Other effects, such as probing different IMF or regions with varying escape fractions of the ionising photons, would change the intrinsic ratio of UV and H$\alpha$.

The IMF would change in the case of stochastic sampling of stellar populations and under-sampling of the high-mass, ionising stars. 
This effect is more dominant in the case of probing small spatial scales (up to $<0.5$ kpc) or areas with $\rm SFR>10^{-4}\, M_\odot/yr$ (\citealt{Lee16}), which is more prevalent when observing  nearby galaxies (\citealt{Calzetti07, Murphy11, Faesi14, Silva14, Krumholz15}).
However, we are probing larger areas ($\approx1$ kpc) with typical SFR values of $\rm >10^{-4}\, M_\odot/yr$. 
Therefore, we conclude that the IMF does not play an important role in $\rm \Delta SFR$ discrepancy.

Higher escape fraction of ionizing photons, due to variable gas and dust density and relative distribution (i.e. patchiness of dust, \citealt{Calzetti94}), could change the intrinsic UV/H$\alpha$ ratios and the relation between the UV slope and the  equivalent width of the Balmer lines (W(H$\alpha$); 
\citealt{Zackrisson13}).
Measuring a proper $\rm W(H_{\alpha})$ in the stripped tails, especially in the DIG regions of our galaxy sample is highly uncertain due to the uncertain measurements and calibration of the stellar component of the observations (\citealt{Tomicic21a}), which makes this approach unfeasible in the context of this work.
Furthermore, proper measuring of the photon escape fraction, and effects of attenuation and dust distribution on the UV slope cannot be properly done in this work due to technical limitations.    

%%%%

%%%%%%%%%%%%%%%%%%%%%%%%%%%
%%%%
%%   Conclusions
%%%%
%%%%%%%%%%%%%%%%%%%%%%%%%%
\section{Summary} \label{sec:Summary}

Measuring the SFRs in galaxies depends on using various tracers of star formation, such as the UV and ionized gas emission (H$\alpha$).
Differences in SFR values from those  tracers, $\rm \Delta SFR\,=\,log_{10}SFR(UV_{corr})-log_{10}SFR(H\alpha_{corr})$, can  indicate changes in the SFR prescriptions due to different physical processes affecting those tracers. 
To observe the gas-stripping process and effects of ram pressure on the ISM of galaxies in galaxy clusters, it is interesting to probe its effects on $\rm \Delta SFR$ across their disks and stripped tails, and in HII and DIG-dominated areas. 
In this work, we investigated spatially resolved  $\rm \Delta SFR$ variations in 5 strongly gas-stripped galaxies (JO201, JO60, JW100, JW39, JO194) from the GASP survey, and expanded the investigation to regions dominated by diffuse ionized gas.
This is the first pixel-by-pixel comparison in UV and H$\alpha$ in the GASP survey (\citealt{Poggianti17}). 
We compared extinction-corrected tracers of H$\alpha$ emission from the ionized gas (MUSE, optical IFU observations) and UV emission from the young stars (UVIT/ASTROSAT telescope) to compare their SFR values.  
This paper explores results at pixel scales of 0.5\,kpc (point-spread function of 1.2-1.4\,kpc resolution)
The observations and our analysis yield the following conclusions:

\begin{itemize}
    \item The regions dominated by dense gas (dominated by star-forming regions) show $\rm \Delta SFR\,\approx0$,  indicating that the SFR prescriptions for UV and H$\alpha$ result in similar SFRs. In contrast, the DIG-dominated regions differ and show $\rm \Delta SFR\approx0.5$  values, with SFR(UV) being higher than SFR(H$\alpha$). This is supporting the scenario according to which in DIG-dominated regions mechanisms different than SF takes place, hence the prescriptions valid in dense-gas-dominated regions are not valid.

    \item There is  a large galaxy-by-galaxy variation in $\rm \Delta SFR$. There is no difference in the $\rm \Delta SFR$ behavior between the disks and the tails. 

    \item We empirically derived the SFR prescriptions for FUV, NUV and H$\alpha$ for these gas-stripped galaxies, for both the disks and tails, and for spatially resolved and integrated cases. We used extinction-corrected H$\alpha$ emission as a reference tracer, from the SF spaxels according to the BPT-[O\textsc{I}] diagram.  We note that only integrated cases were done for H$\alpha$ and that the reference tracer is dominated by high luminosity spaxels. UV prescriptions vary between the galaxies, but the mean values are similar to the literature values within the uncertainty of the data. 
    
    \item The  SFR prescriptions for H$\alpha$ deviate up to 0.2\,dex from the literature values due to a large amount of the diffuse  H$\alpha$  emission in the tails ionized by non-SF sources. This indicates that the ram-pressure process offsets the ISM physical properties with respect to those defining the standard SFR prescription.    

    \item The jellyfish galaxies exhibit a ram-presure front where there are no H$\alpha$ and only UV emission, indicating stripping of the gas from disks to tails. We note that areas  with H$\alpha$ and no UV emission show LINER like features,  with $\rm log_{10}[O\textsc{I}]/H\alpha$ line ratios higher than in SF regions. Those areas are mostly found in the stripped tails. On the other hand, areas with both the UV and H$\alpha$ show predominantly SF source of ionisation in the disk and a mixture of sources in the tails. These results indicate that the gas in the stripped tails is ionised by processes other than SF, potentially from the mixing of the hot ICM gas and cold ISM of the tail.

\end{itemize}

We discussed the potential causes of variations in $\rm \Delta SFR$ between the dense gas and DIG areas, such as changes in escape fractions of ionizing photons, assumptions of attenuation curve, changes in timescales of star formation, changes in the IMF, etc.  
Of those, we conclude that the dominant cause are changes in timescales of star formation. 
The DIG-dominated regions are the ones where active star formation recently ended, while those regions exhibit both the UV emission from stars and  H$\alpha$ from the DIG.

\acknowledgments

Based on observations collected at the European Organization for Astronomical Research in the Southern Hemisphere under ESO program 196.B-0578. 
This project has received funding from the European Research Council (ERC) under the European Union's Horizon 2020 research and innovation programme (grant agreement No. 833824).
We acknowledge funding from the INAF main-stream funding program (PI B. Vulcani).
J. F. acknowledges financial support from the UNAM- DGAPA-PAPIIT IN111620 grant, Mexico.
B. V. and M. G. acknowledge the Italian PRIN-Miur 2017 (PI A. Cimatti).
A. W. acknowledges financial support from ASI through the ASI-INAF agreements 2017-14-H.0 (for papers involving X-ray data).
A. I. acknowledges the INAF founding program 'Ricerca Fondamentale 2022' (PI A. Ignesti).

%%%%%%%%%%%%%%%%%%%%%%%%%%%
%%%%
%%   BIBLIOGRAPHY
%%%%
%%%%%%%%%%%%%%%%%%%%%%%%%%%

\bibliographystyle{aasjournal}
\bibliography{NT_UVHa_paper}

\end{document}